\documentclass[12pt]{article}
\usepackage{amsmath}
\usepackage{amssymb}
\begin{document}
\vskip 2cm
\begin{center}
{\bf {\Large Superspace Unitary Operator in QED with Dirac and \\Complex 
Scalar Fields: Superfield Approach}}\\

\vskip 3.2cm

{\sf D. Shukla$^{(a)}$, T. Bhanja$^{(a)}$, R. P. Malik$^{(a,b)}$}\\
$^{(a)}$ {\it Physics Department, Centre of Advanced Studies,}\\
{\it Banaras Hindu University, Varanasi - 221 005, (U.P.), India}\\

\vskip 0.1cm


\vskip 0.1cm

$^{(b)}$ {\it DST Centre for Interdisciplinary Mathematical Sciences,}\\
{\it Faculty of Science, Banaras Hindu University, Varanasi - 221 005, India}\\
{\small {\sf {E-mails: dheerajkumarshukla@gmail.com; tapobroto.bhanja@gmail.com; 
rpmalik1995@gmail.com}}}

\end{center}

\vskip 3cm

\noindent
{\bf Abstract:} We exploit the strength of the superspace (SUSP) unitary operator to 
obtain the results of the application of the horizontality condition (HC) within the 
framework of augmented version of superfield formalism that is applied to the {\it interacting} 
systems of Abelian  1-form  gauge theories where the $U(1$) Abelian 1-form gauge field couples 
to the Dirac and complex scalar fields in the physical four (3 + 1)-dimensions of spacetime.
These interacting theories are generalized onto a (4, 2)-dimensional supermanifold that 
is parametrized by the four (3 + 1)-dimensional (4D) spacetime variables and a pair of
Grassmannian variables. To derive the (anti-)BRST symmetries for the \textit{matter} fields, 
we impose the gauge invariant restrictions (GIRs) on the superfields defined on the 
(4, 2)-dimensional supermanifold. We discuss various outcomes that emerge out from our 
knowledge of the SUSP unitary operator and its hermitian conjugate. The latter operator
is derived without imposing any operation of hermitian conjugation on the parameters and
fields of our theory from {\it outside}. This is an interesting observation in our present 
investigation.\\

\vskip 1.2cm



\newpage
\noindent
\section{Introduction}

One of the most elegant and geometrically intuitive approaches to the $p$-form ($p = 1,2,3,...$)
gauge theories, described within the framework of Becchi-Rouet-Stora-Tyutin (BRST) formalism,
is the superfield approach (see, e.g. [1-5]). In particular, in refs. [1-3], it has been shown that 
one can derive the proper (i.e. off-shell nilpotent and absolutely anticommuting) (anti-) BRST 
symmetry transformations for the non-Abelian 1-form gauge and corresponding (anti-)ghost fields 
by exploiting the potential and power of the horizontality condition (HC) where the super 
curvature 2-form ($\tilde{F}^{(2)}$), defined on the ($D$, 2)-dimensional supermanifold, is equated 
with the ordinary curvature 2-form ($F^{(2)}$) defined on the $D$-dimensional Minkowskian 
flat spacetime manifold. However, the above superfield formalism [1-5] does {\it not} shed any light on the 
derivation of (anti-)BRST symmetry transformations associated with the matter fields of a given 
interacting non-Abelian 1-form gauge theory where there is a coupling between the gauge field
and the Noether conserved current constructed with the matter fields (as far as the minimal interaction 
in a given gauge theory is concerned).

In a set of papers [6-9], the above superfield formalism [1-5] has been consistently generalized 
so as to derive the proper (anti-)BRST symmetry transformations for the {\it matter} fields 
(in addition to the gauge and (anti-)ghost fields where the input from the outcomes of the HC 
plays an important role (see, e.g. [8,9] for details)).
The generalized version of superfield formalism (where the HC and gauge invariant restrictions
(GIRs) are exploited together) has been christened as the augmented version of superfield formalism. 
In Ref. [1-3], a superspace (SUSP) unitary operator has been intelligently chosen which provides 
the (anti-)BRST symmetry transformations for the matter, (anti-)ghost and gauge fields where the  gauge group 
structure of the specific gauge theory is very elegantly maintained. However, the explicit mathematical 
derivation of this operator has not been provided in these seminal works [1-3]. It would be a nice idea
to exploit the key concepts of augmented superfield formalism to derive this SUSP 
unitary operator clearly.

The purpose of our present paper is to derive the above SUSP unitary operator elegantly and explicitly 
in the case of interacting Abelian 1-from gauge theories with Dirac and complex scalar fields.
In this connection, first of all, we exploit the potential of the HC to derive the (anti-)BRST 
symmetry transformations for the Abelian 1-form gauge and corresponding (anti-)ghost fields. 
Subsequently, we utilize {\it this} result to derive the (anti-)BRST symmetry transformations 
for the matter fields (i.e. Dirac and complex scalar fields) to obtain the explicit form of 
the SUSP unitary operator where the SUSP $U(1)$ gauge group structure is maintained. We exploit 
the explicit mathematical form of this operator to derive the results of HC and prove the reasons
behind the imposition of HC in the superfield approach to BRST formalism. This is one of the highlights 
of our present investigation.

One of the key consequences of the SUSP unitary operator is that the matter field transforms
(e.g. $\Psi^{(g)} (x, \theta, \bar\theta) = U (x, \theta, \bar\theta) \; \psi (x)$) in such
a manner that the SUSP $U(1)$ gauge group structure is respected in the transformation space. As a result,
one can define the covariant derivative which would also transform in exactly the same manner
(i.e. $ \tilde{D} \Psi^{(g)} (x, \theta, \bar\theta) = U (x, \theta, \bar\theta) \,D \psi (x)$).
This, in turn, defines the transformation of the  supercurvature 2-form 
(i.e. $\tilde{F}^{(2)} =  U (x, \theta, \bar\theta) \, F^{(2)} \, U^\dagger (x, \theta, \bar\theta)$) 
which leads to the derivation of the HC (i.e. $ \tilde F^{(2)} = F^{(2)}$) (because the SUSP unitary
operator $U (x, \theta, \bar\theta)$ is Abelian in nature and $U^\dagger \, U = U \, U^\dagger = 1$).
Thus, we obtain an alternative to the HC in the language of the SUSP unitary operator and, in some
sense, we provide the proof for the validity of the HC (i.e. $ \tilde F^{(2)} = F^{(2)}$) in the
context of superfield approach to any arbitrary
$D$-dimensional Abelian gauge theory (described within the framework of BRST formalism).

Our present endeavor is motivated by the following factors. First, the SUSP unitary operator 
$ U(x, \theta, \bar{\theta}) $ has been judiciously chosen in [1-3]. However, it has {\it not} been 
theoretically derived. We have accomplished this goal in our present endeavor. Second, 
the accurate derivation of this SUSP operator provides the {\it proof} behind the imposition of 
the HC in the context of superfield approach to BRST formalism. Third, the $U(1)$ 
group structure appears very naturally in the theory due to the transformation property
(e.g. $\Psi^{(g)} (x, \theta, \bar{\theta}) = U (x, \theta, \bar{\theta})\, \psi (x)$, 
etc.). Fourth, the results of HC are reproduced by using the SUSP unitary operator which provides, 
in some sense, an alternative to it. Finally, our present endeavor for the Abelian theory is our first modest step towards
our main goal of obtaining the SUSP unitary operator for the non-Abelian theory.

Our present paper is organized as follows. In Sec. 2, we discuss the importance of 
HC in the derivation of complete set of proper (anti-)BRST symmetry transformations for 
the gauge and (anti-)ghost fields of this theory. Our Sec. 3 lays emphasis on the 
derivation of (anti-)BRST symmetries for the {\it matter} fields and the SUSP 
unitary operator (which is responsible for the shift transformations along the Grassmannian 
directions of the supermanifold). In Sec. 4, we derive the (anti-)BRST symmetry 
transformations for the gauge and (anti-)ghost fields by exploiting the strength of the 
SUSP unitary operator (which is equivalent to the application of HC). Finally, we make 
some concluding remarks and point out a few future directions for further investigations 
in Sec. 5.

\noindent
\section{Preliminaries: HC and (anti-)BRST symmetries}

We start off with the following (anti-)BRST invariant Lagrangian density $ {\cal L}^{(D)}_B$ for the interacting 
four (3 + 1)-dimensional (4D) 
$U(1)$ gauge theory with Dirac fields $\psi$ and $\bar\psi$ (with mass {\it m} and electric charge 
{\it e}) as\footnote{We adopt here the convention and notations such that the background 4D Minkowskian 
flat spacetime metric $(\eta_{\mu\nu})$ has the signatures $(+1,-1,-1,-1)$ so that $(\partial \cdot A) 
= \partial_\mu A^\mu \equiv \eta_{\mu\nu}\,\partial^\mu A^\nu = \partial_0\, A_0 - \partial_i\,A_i $ 
where the Greek indices $\mu,\, \nu, \, \lambda.. = 0,1,2,3 $ correspond to the spacetime directions 
and the Latin indices $i, j, k... = 1, 2, 3$ stand for the space directions only. }
\begin{eqnarray}
{\cal L}^{(D)}_B &=& - \frac{1}{4}\, F_{\mu\nu}\, F^{\mu\nu} + \bar\psi\,(i \gamma^{\mu} D_{\mu} - m)\, \psi
+ B (\partial \cdot A)  + \frac{B^2}{2}\, -\, i\, \partial_\mu \bar C\, \partial^\mu C,
\end{eqnarray}
where the covariant derivative $ D_{\mu} \, \psi = \partial_\mu \psi + i\,e\,A_\mu \psi$ and the
2-form $F^{(2)} = d\, A^{(1)}$ defines the curvature tensor 
$F_{\mu\nu} = \partial_\mu\, A_\nu - \partial_\nu\, A_\mu$ for the 1-form $A^{(1)} = dx^\mu A_\mu$
connection $A_\mu$ where $d = dx^\mu \partial_\mu$ ( with $d^2 = 0$) is the exterior derivative. In the above,
$B$ field is the Nakanishi-Lautrup auxiliary field which is used for the linearization of the 
gauge-fixing term: [$-\frac{1}{2}(\partial \cdot A)^2$] and $(\bar C)\,C$ are the fermionic
($ C^2 = {\bar C}^2 = 0, C\,\bar C + \bar C\, C = 0$) (anti-)ghost fields. The above
Lagrangian density respects the following (anti-)  BRST symmetry transformations\footnote{We shall use, 
throughout the whole body of our text, the notations $s_{(a)b}$ for the continuous 
and infinitesimal (anti-)BRST symmetry transformations connected with the 4D {\it interacting}
Abelian 1-form gauge theories (of the Dirac and complex scalar fields).} [8] 
\begin{eqnarray}
&& s_b \, A_\mu = \partial_\mu C,\quad\quad \qquad s_b \,C = 0,\qquad \qquad s_b\,\bar{C} = i\, B, \nonumber\\ 
&& s_b B = 0, \quad\qquad s_b\,\psi = -\,i\, e\,C\, \psi, \qquad\qquad s_b\,\bar{\psi} = -\,i\,e\,\bar{\psi}\,C, \nonumber\\
&& s_{ab}\, A_\mu = \partial_\mu \bar{C}, \quad\qquad s_{ab} \,\bar{C} = 0,\qquad \qquad s_{ab}\,C = -\,i\, B, \nonumber\\
&& s_{ab} B = 0, \quad \qquad s_{ab}\,\psi = -\,i\, e\,\bar{C}\, \psi,\quad \qquad s_{ab}\,\bar{\psi} = -\,i\,e\,\bar{\psi}\,\bar{C}.
\end{eqnarray} 
It can be shown that the above transformations are off-shell nilpotent ($s^2_{(a)b} = 0$)
of order two and absolutely anticommuting $(s_b\,s_{ab} + s_{ab}\,s_b = 0)$  in nature.

The 4D (anti-)BRST invariant Lagrangian density ${\cal L}^{(C)}_B$ for the complex scalar fields 
$\varphi (x)$ and $\varphi^* (x)$ (with mass {\it m} and electric charge {\it e}) is (see, e.g. [9])
\begin{eqnarray}
{\cal L}^{(C)}_B &=& -\,\frac{1}{4}\, F^{\mu\nu}F_{\mu\nu} + (D_\mu \varphi)^* \,(D^\mu \varphi) 
- m^2 \varphi^* \varphi + B\,(\partial \cdot A)  + \frac{B^2}{2} - i\,\partial_\mu \bar{C}\,\partial^\mu C,
\end{eqnarray}
where $D_\mu\, \varphi = (\partial_\mu + i\,eA_\mu)\,\varphi$ and $(D_\mu \,\varphi)^* = (\partial_\mu 
- i\,eA_\mu)\, \varphi^*$ are the covariant derivatives on the fields $\varphi (x)$ and $\varphi^* (x)$ 
and the rest of the symbols in ${\cal L}^{(C)}_B$ have been explained after equation (1).
It can be seen that the Lagrangian density ${\cal L}^{(C)}_B $ respects the following off-shell nilpotent 
$(s^2_{(a)b} = 0)$ (anti-)BRST symmetry transformations $s_{(a)b}$, namely;
\begin{eqnarray}
&& s_b \, A_\mu = \partial_\mu C,\quad \qquad s_b \,C = 0,\quad \qquad s_b\,\bar{C} = i\, B, \qquad\quad s_b\, B = 0, \nonumber\\
&& s_b\,\varphi = -\,i\, e\,C\, \varphi, \quad\qquad s_b\,\varphi^* = +\,i\,e\,\varphi^* \,C, \qquad \quad s_{b}\,F_{\mu\nu} = 0,
 \nonumber\\
&& s_{ab}\,A_\mu = \partial_\mu \bar{C},\quad \quad s_{ab} \,\bar{C} = 0,\qquad \quad s_{ab}\,C 
= -\,i\, B, \quad \quad s_{ab}\,B = 0, \nonumber\\ 
&& s_{ab}\,\varphi = -\,i\, e\,\bar{C}\, \varphi, \qquad \quad s_{ab}\,{\varphi}^* 
= +\,i\,e\,{\varphi}^*\,\bar{C},\qquad \quad  s_{ab}\,F_{\mu\nu} = 0.
\end{eqnarray} 
We also note that $s_b$ and $ s_{ab}$ absolutely anticommute $(s_b\,s_{ab} + s_{ab}\,s_b = 0)$ 
with each other. Physically, the nilpotency property encapsulates the fermionic nature of 
(anti-)BRST symmetry transformations and the linear independence of (anti-)BRST symmetry transformations 
is encoded in the property of absolute anticommutativity. It is worthwhile to mention that, 
unlike in the case of {\it fermionic} Dirac fields, the complex scalar fields $\varphi (x)$ and $\varphi^* (x)$ 
{\it commute} with the (anti-)ghost fields $C$ and $\bar{C}$.

To derive the proper (i.e. nilpotent and absolutely anticommuting) (anti-)BRST symmetry 
transformations for the gauge and (anti-)ghost fields, within the framework of superfield 
formalism [1-3], we apply the HC\footnote{Physically, the HC implies that, for the free
Abelian 1-form gauge theory, the gauge invariant physical electric and magnetic fields must 
remain independent of the Grassmannian variables of the (4, 2)-dimensional supermanifold. 
This is an essential requirement from the viewpoint of physics because the Grassmannian 
variables are merely a sort of mathematical artifacts which are useful only in the description 
of superspace formulation. Furthermore,  these variables are {\it not} physically realized  
unlike the spacetime variables.} on the super 1-form ($\tilde{A}^{(1)}$), defined on the 
(4, 2)-dimensional supermanifold (with the help of super exterior derivative $\tilde{d}$) 
as [1-3,8,9]
\begin{eqnarray}
\tilde{d}\,\tilde A^{(1)} = d\, A^{(1)} \quad \Longleftrightarrow\quad \tilde{F}^{(2)} = F^{(2)},
\end{eqnarray}
where $F^{(2)} = \big[(dx^\mu \wedge dx^\nu)\,/\,2 \big]\,F_{\mu \nu}$ is the 
curvature 2-form defined on the 4D ordinary spacetime manifold and $\tilde {F}^{(2)} = 
\big[(dZ^M \wedge dZ^N)\,/\,2 \big]\, \tilde {F}_{MN}$ is the supercurvature 2-form defined  
on the (4, 2)-dimensional supermanifold. We have the following explicit generalizations, namely;
\begin{eqnarray}
&& d = dx^\mu\, \partial_\mu \longrightarrow  \tilde{d} = dZ^M \partial_M = dx^\mu \, \partial_\mu 
+ d\theta\, \partial_\theta + d\bar{\theta}\,\partial_{\bar{\theta}},\nonumber\\
&& A^{(1)} = dx^\mu\, A_{\mu}  \longrightarrow  \tilde{A}^{(1)} = dZ^M A_M 
\equiv   dx^\mu\,B_\mu (x, \theta, \bar{\theta}) 
+ d\theta\, \bar{F}\, (x, \theta, \bar{\theta}) + d\bar{\theta}\, F  (x, \theta, \bar{\theta}),
\end{eqnarray}
where the superspace coordinate $Z^M = (x^\mu, \theta, \bar{\theta})$ and the super derivative 
$\partial_M = (\partial_\mu, \partial_\theta, \partial_{\bar{\theta}})$ characterize the 
(4, 2)-dimensional supermanifold and $A_M = (B_\mu, F, \bar{F})$ corresponds to a vector supermultiplet. 
Here the spacetime coordinates  $x^\mu$ (with $\mu = 0,1,2,3 $) are the bosonic variables and $(\theta, \bar\theta)$ is a 
pair of Grassmannian variables (with $\theta^2 = \bar\theta^2 = 0, \; \theta\,\bar\theta 
+ \bar\theta\,\theta = 0$). The superfields $B_\mu (x, \theta, \bar{\theta}), F (x, \theta, \bar{\theta})$ 
and $\bar{F} (x, \theta, \bar{\theta})$ can be expanded along the Grassmannian directions of the 
(4, 2)-dimensional supermanifold as [1-3,8,9]
\begin{eqnarray}
&& B_\mu (x, \theta, \bar{\theta}) = A_\mu (x) + \theta\, \bar{R_\mu} (x) + \bar{\theta}\, R_\mu (x)
+ i\, \theta\bar{\theta}\,S_\mu (x), \nonumber\\
&&F (x, \theta, \bar{\theta}) = C (x) + i\, \theta\, \bar{B_1} (x) + i\,\bar{\theta}\, B_1 (x) 
+ i\, \theta\bar{\theta} \, s (x), \nonumber\\
&&\bar{F} (x, \theta, \bar{\theta}) = \bar{C} (x) + i\, \theta\, \bar{B_2} (x) + i\, \bar{\theta}\, B_2 (x) 
+ i\,\theta\bar{\theta}\,\bar{s}(x),
\end{eqnarray}
which yield the basic fields $(A_\mu, C, \bar{C})$ of our starting Lagrangian densities (1) and (3) 
in the limit $\theta = \bar{\theta} = 0$. In the above, the fields $(\bar{R}_\mu, R_\mu, S_\mu, 
\bar{B}_1, B_1, s, \bar{s}, \bar{B}_2, B_2)$ are the secondary fields which are to be determined in 
terms of the basic and auxiliary fields of the Lagrangian density (1). In fact, it can be explicitly 
checked that we obtain (see, e.g. [8,9] for details)
\begin{eqnarray}
&& R_\mu = \partial_\mu C, \qquad  S_\mu = \partial_\mu B, \qquad  \bar{R}_\mu = \partial_\mu \bar{C},  \quad
B_1 = s = \bar{s} = \bar{B}_2 = 0, \nonumber\\
&& \bar{B}_1 + B_2 = 0 \qquad \Longrightarrow \quad \bar{B}_1 = -\, B = -\,B_2,
\end{eqnarray}
when we exploit the HC\footnote{The horizontality condition {\it physically}
implies that the electric and magnetic fields of the Abelian Maxwell's theory 
should be independent of the presence of the Grassmannian variables in SUSY theory.} (5). The substitution of (8) into (7) yields 
\begin{eqnarray}
 B^{(h)}_\mu (x, \theta, \bar{\theta}) &=& A_\mu  + \theta\, (\partial_\mu \bar{C}) 
+ \bar{\theta}\, (\partial_\mu C) + \theta\bar{\theta}\, (i\,\partial_\mu B) \nonumber\\
 &\equiv&  A_\mu  + \theta\, (s_{ab}\, A_\mu) + \bar{\theta}\,(s_b \, A_\mu) 
+ \theta\bar{\theta}\, (s_b\, s_{ab}\, A_\mu),\nonumber\\
 F^{(h)} (x, \theta, \bar{\theta}) &=& C + \theta\, (-\, i\,B) \equiv C + \theta \, (s_{ab}\,C), \nonumber\\
 \bar{F}^{(h)} (x, \theta, \bar{\theta}) &=& \bar{C} + \bar{\theta}\, (i\,B) \quad \equiv \bar{C}  
+ \bar{\theta}\, (s_b \,\bar{C}),
\end{eqnarray}
where the superscript $(h)$ stands for the superfields that have been derived after the application 
of HC. It is clear, from the above, that we have already obtained the (anti-) BRST symmetry transformations 
for the gauge and (anti-)ghost fields (cf. (2) and (4)) for the interacting system of the 
$U(1)$ Abelian 1-form gauge theories (with Dirac and complex scalar fields). It is to be noted that the 
(anti-)BRST transformations for the gauge and (anti-)ghost fields are the {\it same} for both the 
interacting $U(1)$ gauge theories under consideration. Furthermore, it is interesting to point out 
that the (anti-)BRST symmetry transformations ($s_{(a)b}$) are connected with the translational 
generators $\partial_\theta $ and $\partial_{\bar\theta} $ along the Grassmannian directions of 
the supermanifold by the relationships: $s_b \Longleftrightarrow \partial_{\bar{\theta}},\;
s_{ab} \Longleftrightarrow \partial_\theta  $.

\noindent
\section{Gauge invariant restrictions and SUSP unitary operator: (anti-)BRST symmetries for matter fields}

In our previous section, we have derived the (anti-)BRST symmetry transformations for the 
gauge and (anti-)ghost fields but have not discussed the (anti-)BRST symmetries associated 
with the Dirac fields. To obtain these symmetry transfromations, we impose the following gauge invariant
restriction (GIR) on the matter superfields (see, e.g. [8] for details)
\begin{eqnarray}
\bar\Psi (x, \theta, \bar{\theta})\, \tilde D (x, \theta, \bar{\theta})\, \Psi (x, \theta, \bar{\theta})
= \bar\psi (x) \, D\, \psi (x),
\end{eqnarray}
where $D = d + i\, e\, A^{(1)}$ and  $\tilde D = \tilde d + i\, e\, \tilde A^{(1)}_{(h)}$.
The super 1-form $ \tilde A^{(1)}_{(h)}$ connection 
(with $\tilde d \tilde A^{(1)}_{(h)} = d A^{(1)}$) is defined, in terms of the superfields (9), as follows
 \begin{eqnarray}
\tilde A^{(1)}_{(h)}(x, \theta, \bar{\theta}) = dx^\mu\,B^{(h)}_\mu (x, \theta, \bar{\theta}) 
+ d\theta\, \bar{F}^{(h)}\, (x, \theta, \bar{\theta}) + d\bar{\theta}\, F^{(h)}  (x, \theta, \bar{\theta}), 
\end{eqnarray}
where the explicit expansions of $B^{(h)}_\mu (x, \theta, \bar{\theta}), 
\bar{F}^{(h)}\, (x, \theta, \bar{\theta})$ and $F^{(h)} (x, \theta, \bar{\theta})$ are 
given in (9). The matter superfields $ \bar\Psi (x, \theta, \bar{\theta})$ and 
$ \Psi (x, \theta, \bar{\theta})$ have the following expansions along the Grassmannian 
directions of the (4, 2)-dimensional supermanifold, namely; 
\begin{eqnarray}
&&\Psi (x, \theta, \bar{\theta}) = \psi(x) + i\,\theta\, \bar b_1 (x) + i\,\bar\theta\,b_1(x) + i\,\theta\,\bar\theta\, t(x),\nonumber\\
&&\bar\Psi (x, \theta, \bar{\theta}) = \bar\psi(x) + i\,\theta\, \bar b_2 (x) + i\,\bar\theta\,b_2(x)
 + i\,\theta\,\bar\theta\, \bar t(x),
\end{eqnarray}
where ($b_1, \bar b_1, b_2, \bar b_2, t, \bar t $) are the secondary fields which
would be determined in terms of the basic and auxiliary fields of our present theory
described by the Lagrangian density (1). In this connection, the GIR in (10) helps us 
to obtain the following relationships between the secondary fields and basic and
auxiliary fields\footnote{We slightly differ with the relationships
mentioned in [8] because there is a minor printing error in the latter  as far as the relationships
quoted for $t$ and $\bar t$ are concerned. The relationships in (13) are {\it all} correct.} (see, e.g. [8] for details)
\begin{eqnarray}
&& \bar b_1 = - e\, \bar C\, \psi, \quad  b_1 = - e\, C\, \psi, \quad t 
= -\,i\,e\,(B - e\,C\,\bar C),\nonumber\\
&& \bar b_2 = - e\,\bar \psi\, \bar C, \quad  b_2 = - e\, \bar\psi\,C, 
\quad \bar{t} = i\,e\,(B - e\,\bar C\, C).
\end{eqnarray} 
The substitution of these expressions into (12) yields the following explicit expansions for the
matter superfields in terms of the (anti-)BRST symmetry transformations (2):
\begin{eqnarray}
\Psi^{(g)} (x, \theta, \bar{\theta}) &=& \psi(x) + \theta\, (- \,i\, e\, \bar C\, \psi) 
+ \bar\theta\,(-\,i\, e\, C\, \psi) 
+ \theta\,\bar\theta\, (e\,(B - e\,C\,\bar C)\, \psi)\nonumber\\
&\equiv & \psi(x) + \theta\,(s_{ab}\, \psi) + \bar\theta \, (s_{b}\, \psi) + \theta\,\bar\theta (s_b\,s_{ab}\, \psi),\nonumber\\
\bar\Psi^{(g)} (x, \theta, \bar{\theta}) &=&  \bar\psi(x) + \theta\, (- \,i\, e\, \bar \psi\, \bar C) 
+ \bar\theta\,(-\,i\, e\, \bar\psi\,C) 
+ \theta\,\bar\theta\, (-\, e\, \bar \psi\, (B - e\,\bar C\, C))\nonumber\\
 &\equiv & \bar\psi(x) + \theta\,(s_{ab}\, \bar\psi) + \bar\theta \, (s_{b}\, \bar\psi)
+ \theta\,\bar\theta (s_b\,s_{ab}\, \bar\psi),
\end{eqnarray}
where the superscript $(g)$ denotes the expansions of the superfields obtained after the 
application of the GIR  (10).

We are now in the position to state the precise form of the SUSP unitary operator which transforms
the ordinary Dirac matter fields $\psi (x)$ and $\bar\psi (x)$ to their counterparts
$\Psi^{(g)} (x, \theta, \bar{\theta})$ and $\bar\Psi^{(g)} (x, \theta, \bar{\theta})$. In fact,
using the expansions (14), it is clear that\footnote{We note that the relationship 
$\Psi^{(g)} (x, \theta, \bar{\theta}) = U(x, \theta, \bar{\theta})\,\psi (x) $ is exactly of 
the same kind as the $U(1)$ gauge transformation on the Dirac field: 
$\psi (x) \longrightarrow \psi' (x) = U (x) \,\psi (x)$ where the operator $U(x) = e^{-\,i\,e\,\alpha (x)}$
(with gauge parameter $\alpha (x)$) forms the $U(1)$ group as it satisfies {\it all} the group properties 
under product.} 
\begin{eqnarray}
\Psi^{(g)} (x, \theta, \bar{\theta}) &=& [ 1 +\theta\, (- \,i\, e\, \bar C) 
+ \bar\theta\,(-\,i\, e\, C) 
+ \theta\,\bar\theta\, \left( e\,(B - e\,C\,\bar C ) \right)]\; \psi(x) \nonumber\\
&\equiv & U(x, \theta,\bar\theta)\, \psi (x),\nonumber\\
\bar\Psi^{(g)} (x, \theta, \bar{\theta}) &=&  \bar\psi(x)\,[ 1 + \theta\, (i\, e\, \bar C) 
+ \bar\theta\,(i\, e\,C)  
+ \theta\,\bar\theta\, \left(-\, e\,(B - e\,\bar C\, C)\right)]\nonumber\\
&\equiv & \bar\psi(x)\,U^\dagger(x, \theta,\bar\theta), 
\end{eqnarray}
where $U(x, \theta,\bar\theta)$ and $U^\dagger(x, \theta,\bar\theta)$ are the SUSY generators which,
primarily, lead to the shift transformations along the Grassmannian directions (because $\Psi^{(g)} 
(x, \theta, \bar{\theta}) = U(x, \theta,\bar\theta)\, \psi (x)$ and $\bar\Psi^{(g)} (x, \theta, \bar{\theta}) 
= \bar\psi(x)\,U^\dagger(x, \theta,\bar\theta)$). These SUSP operators (i.e. $U$ and $U^\dagger$)
can be expressed in the mathematically precise exponential forms as\footnote{Under the hermitian conjugation
operations: $\theta^\dagger = \mp \;\theta, \bar\theta^\dagger = \mp \;\bar\theta, C^\dagger = \pm \;C, \bar C^\dagger 
= \pm \;\bar C, B^\dagger = B, e^\dagger = e, i^\dagger = - i$, it can be readily checked that SUSP
operators $U (x, \theta, \bar\theta)$ and $U^\dagger (x, \theta, \bar\theta)$ interchange with each-other and the FP-ghost part 
(i.e. $- \,i \,\partial_\mu \bar C \,\partial^\mu C $) of the Lagrangian densities (1) and (3) remains invariant.} 
\begin{eqnarray}
 U(x, \theta,\bar\theta) &=& exp\, \bigl [ \theta\, (- \,i\, e\, \bar C) + \bar\theta\,(-\,i\, e\, C) 
+ \theta\,\bar\theta\, \left(e\,B\right) \bigr ], \nonumber\\
 U^\dagger(x, \theta,\bar\theta) &=& exp \, \bigl [ \theta\, (i\, e\,  \bar C) 
+ \bar\theta\,(i\, e\, C) +  \theta\,\bar\theta\, \left(-\, e\,B \right) \bigr ],
\end{eqnarray}
which directly establish that the SUSP operator $U$ is unitary
(i.e.  $U\, U^\dagger = U^\dagger\,U =  1)$. This statement can be proven to be true
by using the explicit expressions for $U$ and $U^\dagger$ that are 
quoted in (15) (and that are equivalent to (16)). The crucial observation is that the SUSP operator
$ U(x, \theta,\bar\theta)$ forms a $U(1)$ group in the space of transformations 
where the exponential form (16) of the operator $U(x, \theta,\bar\theta)$
plays an important role. Similar statement could be made with the operator $U^\dagger (x, \theta,\bar\theta)$, too.

To obtain the (anti-)BRST symmetry transformations associated with the complex scalar fields 
$\varphi (x)$ and $\varphi^* (x)$ (c.f. Eq. (4)), we impose the following gauge invariant restrictions
(GIRs) on the superfields defined in the (4, 2)-dimensional supermanifold [9]
\begin{eqnarray}
&& \Phi^\star (x, \theta, \bar\theta)\, \left(\tilde{d} + i\,e\tilde{A}^{(1)}_{(h)}\right)\, \Phi (x, \theta, \bar\theta) 
 = \varphi^* (x)\,\left(d + i\,e A^{(1)}\right)\,\varphi (x), \nonumber\\
&& \Phi (x, \theta, \bar\theta)\,\left(\tilde{d} - i\,e\tilde{A}^{(1)}_{(h)}\right)\,\Phi^\star (x, \theta, \bar\theta) 
 = \varphi (x) \,\left(d - i\,e A^{(1)}\right)\,\varphi^* (x),
\end{eqnarray}
where the superfields $\Phi (x, \theta, \bar\theta)$ and $\Phi^\star (x, \theta, \bar\theta)$ have the 
expansions along the Grassmannian directions of the (4, 2)-dimensional supermanifold as [9]
\begin{eqnarray}
\Phi(x, \theta, \bar\theta) &=& \varphi (x) + i\,\theta \bar{f}_1 (x) + i\, \bar\theta f_2 (x) 
+ i\, \theta \bar\theta \, b (x), \nonumber\\
\Phi^\star(x, \theta, \bar\theta) &=& \varphi^* (x) + i\theta {f}^*_2 (x) + i \bar\theta f^*_1 (x) 
+ i \theta \bar\theta b^*(x).
\end{eqnarray}
In the above, we have the secondary fields on the r.h.s.
as $(\bar{f}_1, f^*_1, f_2, f^*_2, b, b^*)$. These fields could be determined in terms of the basic 
and auxiliary fields of the Lagrangian density ${\cal L}^{(C)}_B$ 
due to the GIRs in (17). It is worthwhile to mention that the r.h.s. of (17) are gauge invariant quantities
and, therefore, they are (anti-)BRST invariant, too.

In our earlier work [9], all the secondary fields of (18) have been determined in a systematic 
manner by exploiting the strength of GIRs in (17). The outcome is:
\begin{eqnarray}
&& \bar{f}_1 = -\,e\, \bar{C}\varphi,\qquad f_2 = -\,e\,C \varphi,\qquad b = -\,i\,e \,(B - e\,C\, \bar C)\,\varphi, \nonumber\\
&& f^*_1 = e\,C\,\varphi^*, \qquad  f^*_2 = e\,\bar{C}\,\varphi^*, \qquad \,\,\, b^* = i\,e \,(B - e\,\bar{C}C)\,\varphi^*.
\end{eqnarray}
The substitution of these expressions into the expansion (18) leads to the following explicit expansions 
in terms of the (anti-)BRST symmetry transformations (4), namely;
\begin{eqnarray}
 \Phi^{(g)}(x, \theta, \bar\theta) &=& \varphi (x) + \theta\, (-i\,e\, \bar{C}\, \varphi) 
+ \bar\theta \,(-i\,e\,C\,\varphi)  + \, \theta \bar\theta \,\left[e(B - e\, C\bar{C})\, \varphi\right]\nonumber\\
& \equiv & \varphi (x) + \theta\,(s_{ab}\varphi) + \bar\theta \,(s_b \, \varphi) + \theta \bar{\theta}\, (s_b\,s_{ab}\, \varphi), \nonumber\\
\Phi^{\star{(g)}} (x, \theta, \bar\theta) &=& \varphi^* (x) + \theta \,(i\,e\, \varphi^* \bar{C}) 
+ \bar\theta \,(i\,e \,\varphi^*\, C) 
 + \,\theta \bar\theta\, \varphi^* (x) \left[-\,e\, (B - e\,\bar{C}C)\right] \nonumber\\
 &\equiv & \varphi^* (x) + \theta\,(s_{ab}\varphi^*) + \bar\theta \,(s_b \, \varphi^*) 
+ \theta \bar{\theta}\, (s_b\,s_{ab}\, \varphi^*),
\end{eqnarray}
where the superscript $(g)$ denotes the expansions of the superfields after the application of GIRs in (17).
It is pretty obvious that the above superfields can be expressed in terms of the SUSP unitary operators
$U (x, \theta, \bar{\theta})$ and $U^\dagger (x, \theta, \bar\theta)$ exactly like (15) and (16) where
$\Psi^{(g)} (x, \theta, \bar{\theta}) $ and $\bar{\Psi}^{(g)} (x, \theta, \bar{\theta}) $ would be replaced 
by the superfields $\Phi^{(g)} (x, \theta, \bar{\theta})$ and $\Phi^{\star (g)} (x, \theta, \bar{\theta})$ 
of (20). Thus, we note that the form of the SUSP unitary operators (16) remains the {\it same} for both {\it interacting} 
models of QED where there is an interaction between the $U(1)$ gauge field $A_\mu$ and 
the Noether conserved current constructed by the Dirac fields as well
as the charged complex scalar fields. This happens because of the 
existence of the local $U(1)$ gauge group behind the construction of both 
these interacting theories.

\noindent
\section{SUSP unitary operator and HC: salient features}

As a result of the group property in the transformation space, we can define a super 
covariant derivative on the Dirac superfield in the following fashion
\begin{eqnarray}
 \psi (x) &\rightarrow& \Psi (x, \theta, \bar{\theta}) 
= {U} (x, \theta, \bar{\theta})\, \psi(x), \nonumber\\
 D\psi (x) &\rightarrow & \tilde{D}\Psi (x, \theta, \bar{\theta}) 
= {U} (x, \theta,\bar{\theta})\, D\psi (x),
\end{eqnarray}
where $\tilde{D} = \tilde{d} + i\,e \tilde{A}^{(1)}_{(h)}(x, \theta, \bar{\theta})$ 
and $D =  d + i\,e A^{(1)} (x)$. It is now crystal clear that the super 1-form connection 
$\tilde{A}^{(1)}_{(h)} (x, \theta, \bar{\theta})$ and the ordinary 1-form connection 
$A^{(1)}(x)$ are connected by the following equation due to the relationships quoted in 
(21), namely;
\begin{eqnarray}
\tilde{A}^{(1)}_{(h)} (x, \theta, \bar{\theta}) &=& U (x, \theta, \bar{\theta}) 
A^{(1)} (x)\, U^\dagger(x, \theta, \bar{\theta})
+ \frac{i}{e} \, \left(\tilde{d}U (x, \theta, \bar{\theta})\right)\,
U^\dagger (x, \theta, \bar{\theta}) \nonumber\\
& \equiv &  dx^\mu\,B^{(h)}_\mu (x, \theta, \bar{\theta}) +\, d\theta \,
\bar{F}^{(h)}(x, \theta, \bar{\theta}) 
+\, d\bar\theta\, F^{(h)} (x, \theta, \bar{\theta}).
\end{eqnarray}
It is evident that if we use the Abelian $U(1)$ nature of the operator 
$\hat{U}(x, \theta, \bar{\theta})$, the first term on the r.h.s. of (22) yields the following 
expression
\begin{eqnarray}
U (x, \theta, \bar{\theta})\, A^{(1)}(x)\, U^\dagger (x, \theta, \bar{\theta}) 
= A^{(1)} (x) \equiv dx^\mu A_\mu (x).
\end{eqnarray}
The second term on the r.h.s. [i.e. $+ (\frac{i}{e})\,(\tilde{d}\, U)\, U^\dagger $)] leads to
the following explicit expression, namely; 
\begin{eqnarray}
 dx^\mu \left[ \theta\, (\partial_\mu \bar C) + \bar\theta \,(\partial_\mu {C}) 
+ \theta \,\bar{\theta}\, (i\,\partial_\mu B) \right] 
 + \,d\theta\, \left[\bar{C} (x) + i\, \bar\theta \,B(x)\right] + d\bar\theta \, \left[C(x) 
- i\, \theta B(x) \right],
\end{eqnarray}
where we have used the expansions for $\tilde{d}, U (x, \theta, \bar{\theta})$ and 
$U^\dagger (x, \theta, \bar{\theta})$ from equations (6) and (15). Now, it is obvious that 
the comparison of the coefficients of $dx^\mu, d\theta $ and $d\bar\theta $ leads to the 
derivation of  (anti-)BRST symmetry transformations {\it exactly} in the same manner as has 
been done in Sec. 2 where we have exploited the HC. In other words, we obtain exactly 
the same expressions for the $B^{(h)}_\mu, F^{(h)}$ and $\bar{F}^{(h)}$ as has been defined 
in Eq. (9).

From the above discussion, we can claim that the HC used in Eq. (5) is equivalent to the 
relationship (22) where the SUSP unitary operator plays a decisive role. To corroborate the 
above claim, we note that the following property of the ordinary covariant derivatives on the 
Dirac field is true, namely;
\begin{eqnarray}
DD\,\psi (x) = i\,e\, {F}^{(2)}\, \psi (x), 
\end{eqnarray}
where the covariant derivative $D =  dx^\mu (\partial_\mu + i\,e A_\mu)$ and $F^{(2)} 
= \left[(dx^\mu \wedge dx^\nu)/2\right]F_{\mu\nu}(x)$.
This property can be expressed in terms of the SUSP unitary operator as:
\begin{eqnarray}
DD\,\psi (x) \longrightarrow \tilde{D}\tilde{D} \, \Psi (x, \theta, \bar{\theta})
= i\,e\tilde{F}^{(2)} \,U (x, \theta, \bar{\theta})\, \psi (x),
\end{eqnarray}
where $D, \tilde{D}$ and $\tilde{F}^{(2)}$ are defined earlier. Now, using the relationship 
given in (21), we obtain $\tilde{D} =  U\, D\, U^\dagger $. If we substitute this value into 
the l.h.s. (i.e. $ \tilde{D}\tilde{D}\,\Psi $) of (26), we obtain the following relationship
\begin{eqnarray}
 U\,DD\,\psi = i\,e \,\tilde{F}^{(2)}\,U\psi 
 \Longrightarrow  \tilde{F}^{(2)} 
=  U (x, \theta, \bar{\theta})\,F^{(2)} (x) \,U^\dagger (x, \theta, \bar{\theta}).
\end{eqnarray}
Focusing on the Abelian nature of $U (x, \theta, \bar{\theta})$ 
and $U^\dagger (x, \theta, \bar{\theta})$, it is crystal clear that the r.h.s of (27) 
would yield $F^{(2)}$ only (because $U\,U^\dagger = U^\dagger U = I$). Thus, we have 
obtained the HC condition (5) for the Abelian theory (i.e. 
$\tilde{F}^{(2)} = F^{(2)}\,\Longleftrightarrow \; \tilde{d}\,\tilde{A}^{(1)}_{(h)} = d\,A^{(1)}$).

The above argument and discussion can be replicated in the context of QED with complex scalar fields 
where, once again, we obtain the analogue of relation (22) which provides an alternative to the HC. 
Furthermore, we note that the following is true in the context of QED with complex scalar fields, namely;
\begin{eqnarray}
DD\, \varphi = i\,e\,F^{(2)}\, \varphi, \qquad \qquad (DD \varphi)^* = -\,i\,e\, F^{(2)}\, \varphi^*.
\end{eqnarray} 
The above equation can be translated into the superfield formalism as
\begin{eqnarray}
&& DD \, \varphi (x) \rightarrow  \tilde{D}\tilde{D}\, \Phi (x, \theta, \bar{\theta})
= i\,e\, \tilde{F}^{(2)}\, \Phi (x, \theta, \bar{\theta}) 
 \equiv i\,e\,\tilde{F}^{(2)}\, U (x, \theta, \bar{\theta})\,\varphi (x) \nonumber\\
&& (DD \, \varphi)^*   \rightarrow  (\tilde{D}\tilde{D}\, \Phi)^\star (x, \theta, \bar{\theta})
= -\,i\,e\, \tilde{F}^{(2)}\, \Phi^\star (x, \theta, \bar{\theta})
 \equiv   -\,i\,e\, \tilde{F}^{(2)}\, \varphi^* (x)\, U^\dagger (x, \theta, \bar{\theta}),
\end{eqnarray}
where we have: $\tilde{D}\,\Phi (x, \theta, \bar{\theta}) = \left( \tilde{d} 
+ i\,e\,\tilde{A}^{(1)}_{(h)}\right)\,\Phi  (x, \theta, \bar{\theta})$ and 
$(\tilde{D}\,\Phi)^\star (x, \theta, \bar{\theta}) = \left( \tilde{d} 
- i\,e\,\tilde{A}^{(1)}_{(h)}\right)\,\Phi^\star (x, \theta, \bar{\theta})$.
With the input $\tilde{D} = U\,D\,U^\dagger$, it can be checked that
\begin{eqnarray}
 U (x, \theta, \bar{\theta})\,D\,D \,\varphi = i\,e\,\tilde{F}^{(2)}\, 
\Phi (x, \theta, \bar{\theta}) 
\Longrightarrow  \tilde{F}^{(2)} = U (x, \theta, \bar{\theta}) 
\,F^{(2)}\,U^\dagger (x, \theta, \bar{\theta}).
\end{eqnarray} 
The Abelian nature of $U (x, \theta, \bar{\theta})$, once again, implies that we have obtained 
an alternative to the HC (i.e. $\tilde{F}^{(2)} = F^{(2)}$) in the language of the SUSP unitary 
operator $U(x, \theta, \bar{\theta})$ because we have $ U (x, \theta, \bar{\theta}) 
\,F^{(2)} (x) \,U^\dagger (x, \theta, \bar{\theta})  = F^{(2)} (x)$ on the r.h.s. of (30).

The celebrated HC can also be obtained from the relationship given in (22). This is due to
the fact that when we operate by $\tilde d$ on this equation from the left, we obtain the following
explicit equation, namely;
\begin{eqnarray}
\tilde d \, \tilde{A}^{(1)}_{(h)} (x, \theta, \bar{\theta}) 
= \tilde d\, A^{(1)} (x) - \frac{i}{e} \, \tilde{d} \, U (x, \theta, \bar{\theta})\,
\tilde d \,U^\dagger (x, \theta, \bar{\theta}),
\end{eqnarray} 
where we have used (23) and the property $\tilde d^2 = 0$. The first term on the r.h.s. of the above equation
produces: $ \tilde d A^{(1)} = d A^{(1)} = F^{(2)}$ due to the fact that 
the ordinary 1-form ($A^{(1)} (x)$)  connection is independent of
the Grassmannian variables implying that $\partial_\theta A^{(1)} = \partial_{\bar\theta} A^{(1)} = 0$
when we use the definition of $\tilde d$ from (6).
Taking the explicit expressions  for the $U (x, \theta, \bar{\theta}) $ and $ U^\dagger (x, \theta, \bar{\theta})$
from (15), it can be checked that $\tilde d\, U\, \tilde d \,U^\dagger = 0$. This statement becomes very clear
if we take a close look at the exponential forms of $U$ and $U^\dagger$ in (16). It is obvious that the quantity
in the exponent of these operators differ only by a sign factor. Thus, it can be readily checked that\footnote{ Using (15),
it is obvious that $ \tilde d\, U = d x^\mu [ - i \,e \,\theta \,\partial_\mu \bar C 
- i \,e \,\bar\theta \,\partial_\mu C 
+\, e \,\theta \,\bar \theta \,\partial_\mu ( B - e \,C \,\bar C) ] 
+ \,d \theta \,[ - i\,e\, \bar C + e \, \bar\theta\, (B - e \,C \,\bar C) ]
+ \,d \bar\theta\, [ - i\,e\,  C - e \, \theta\, (B - e \,C \,\bar C) ] $ and
$ \tilde d\, U^\dagger = d x^\nu [ + i \,e \,\theta \,\partial_\nu \bar C  
+\, i \,e \,\bar\theta \,\partial_\nu C - \,e \,\theta \,\bar \theta \,\partial_\nu ( B - e \,\bar C \,C) ] 
+\, d \theta \,[ + i\,e\, \bar C - e \, \bar\theta\, (B - e \,\bar C \, C) ] 
+ \,d \bar\theta \, [ + i\,e\,  C + e \, \theta\, (B - e \,\bar C \, C) ]$. From these lucid expressions, one can {\it also} check
that $\tilde d\, U\, \tilde d \,U^\dagger = 0$.}
$\tilde d\, U\, \tilde d \,U^\dagger = 0$.  
This implies that the last term in (31) is zero
(i.e.  $\tilde d\, U\, \tilde d \,U^\dagger = 0$) which, ultimately, leads to the validity of 
HC (i.e. $\tilde F^{(2)} = F^{(2)}$). For readers' convenience, 
we have carried out the explicit computations 
of $\tilde d U$ and $\tilde d U^\dagger$ which are present now in the footnote number 6
and it can be re-checked that the second term on the r.h.s. of (31) is zero (i.e. $\tilde d\, U\, \tilde d \,U^\dagger = 0$).

\noindent
\section{Conclusions}

The central objective of our present investigation has been to derive an explicit expression for 
the SUSP unitary operator $U (x, \theta, \bar{\theta})$ which generates the shift symmetry 
transformations on the gauge, (anti-)ghost and matter superfields along the Grassmannian directions 
of the (4, 2)-dimensional supermanifold on which the ordinary 4D interacting Abelian 1-form 
gauge theories (with Dirac and complex scalar fields) are generalized. In fact, the SUSP unitary operator,
ultimately, leads to the derivation of proper (i.e. off-shell nilpotent and absolutely anticommuting)
(anti-)BRST symmetry transformations for the above {\it interacting} Abelian 1-form gauge theories in
physical four (3 + 1)-dimensions of spacetime.

One of the highlights of our present endeavor is the observation that the correct derivation of the SUSP 
unitary operator provides an alternative to the HC in addition to encompassing in its folds the sanctity of
the $U (1)$ gauge group structure in the transformation space. It is the latter property which allows us to 
define the covariant derivative on the super matter fields (cf. (21) and (29)). This definition, in turn, 
leads to the derivation of a connection between the supercurvature 2-form and the ordinary curvature 2-form 
(cf. (26) and (29)). This is due to the fact that the commutator of two covariant derivatives defines
the field strength tensor $F_{\mu\nu}$ through the relationship $\left[D_\mu, D_\nu \right] \,\psi
= i\,e\,F_{\mu\nu}\,\psi$. This result has been captured in the relationships given in 
eqs. (25), (26) and (28). One of the key features of our present endeavor is the observation
that $U (x, \theta, \bar\theta) $ and $U^\dagger (x, \theta, \bar\theta)$, 
for both the interacting theories, turn out to be the {\it same}.

It would be a nice future endeavor to extend our present idea to derive explicitly 
the SUSP unitary operator in the context of {\it interacting} 4D non-Abelian 1-form 
gauge theory with Dirac fields which has been intelligently chosen in [1-3]. We also 
plan to pursue this direction of investigation in the context of {\it interacting} 
higher $p$-form ($p = 2, 3..$)  gauge theories which are the limiting cases of 
(super)string theories (see, e.g. [10]). We are presently intensively involved with these issues and our
future publications would resolve these in a cogent and convincing manner [11].

\vskip 1cm

\noindent
{\bf Acknowledgments}\\\\
OOne of us (RPM) would like to express his deep sense of gratitude to the AS-ICTP, Trieste, Italy 
and SISSA, Trieste, Italy for the warm hospitality extended to him during his participation in the 
conference on ``Aspects of Gauge and String Theories" (1 - 2 July 2015) which was held at SISSA to
celebrate the $70^{th}$ birth anniversary of L. Bonora. The idea behind our present work came
during this conference. D. Shukla is thankful to the UGC, Government 
of India, New Delhi, for financial support through RFSMS-SRF scheme and T. Bhanja is grateful to 
the BHU-fellowship under which the present investigation has been carried out. Last but not the least,
very useful, sharp and enlightening comments by the Reviewers are gratefully acknowledged, too.

\end{document}